\documentclass[sn-nature, iicol]{sn-jnl}% Style for submissions to Nature Portfolio journals

\usepackage{graphicx}%
\usepackage{multirow}%
\usepackage{amsmath,amssymb,amsfonts}%
\usepackage{amsthm}%
\usepackage{mathrsfs}%
\usepackage[title]{appendix}%
\usepackage{xcolor}%
\usepackage{textcomp}%
\usepackage{manyfoot}%
\usepackage{booktabs}%
\usepackage{algorithm}%
\usepackage{algorithmicx}%
\usepackage{algpseudocode}%
\usepackage{listings}%
\theoremstyle{thmstyleone}%

\theoremstyle{thmstyletwo}%

\theoremstyle{thmstylethree}%

\begin{document}

\title[Article Title]{A Deep Learning-based Compression and Classification Technique for Whole Slide Histopathology Images}

\author[1]{\fnm{Agnes} \sur{Barsi}}\email{ab2076@pgr.aru.ac.uk}

\author[2]{\fnm{Suvendu Chandan} \sur{Nayak}}\email{suvendu2006@gmail.com}

\author[2]{\fnm{Sasmita} \sur{Parida}}\email{sasmitamohanty5@gmail.com}
\author*[1]{\fnm{Raj Mani} \sur{Shukla}}\email{raj.shukla@aru.ac.uk}

\affil*[1]{\orgdiv{Computing and Information Science}, \orgname{Anglia Ruskin University, Cambridge}}
%\orgaddress{\street{Street}, \city{City}, \postcode{100190}, \state{State}, \country{Country}}}

\affil[2]{\orgdiv{Department of Computer Science and Engineering}, \orgname{Silicon Institute of Technology, Bhubaneshwar}}
%\orgaddress{\street{Street}, \city{City}, \postcode{10587}, \state{State}, \country{Country}}}

%%==================================%%
%% Abstract %
%%==================================%%

\abstract{This paper presents an autoencoder-based neural network architecture to compress histopathological images while retaining the denser and more meaningful representation of the original images.  Current research into improving compression algorithms is focused on methods allowing lower compression rates for Regions of Interest (ROI-based approaches). Neural networks are great at extracting meaningful semantic representations from images, therefore are able to select the regions to be considered of interest for the compression process. In this work, we focus on the compression of whole slide histopathology images. The objective is to build an ensemble of neural networks that enables a compressive autoencoder in a supervised fashion to retain a denser and more meaningful representation of the input histology images. Our proposed system is a simple and novel method to supervise compressive neural networks. We test the compressed images using transfer learning-based classifiers and show that they provide promising accuracy and classification performance. }

%\keywords{Image compression, Histopathological image, Encoder, Neural network, Supervised learning}



\maketitle

\section{Introduction}
\label{introduction}

Computers and digital technologies are taking over the world of medicine. This is especially apparent in the field of histopathology which focuses on the microscopic analysis of tissue samples to provide diagnostic insights. Numerous healthcare providers, companies, and startups focus on delivering diagnostic services using state-of-the-art AI solutions \cite{fu2021model}. But these technologies are notoriously data-hungry. The transmission and storage of high-resolution histopathology images are laborious and quite costly. The need to improve the current image compression algorithms that would enable these images to be compressed to satisfying levels without impairing their diagnostic value is ever-increasing \cite{hussain2018image}.

It is no longer necessary to be physically near the microscope or the site at which the tissue sample is taken to analyse it and provide a diagnosis for the patient. Moreover, these digital Whole Slide Images (WSIs) can also be transferred to a remote location through the internet for further requirements. The evolution of digital pathology is firmly underpinned by the development and improvement of Internet technology and cloud architectures in healthcare. Having sufficient broadband connections to transmit WSI images for telepathology is just scratching the surface \cite{Griffin2017DigitalPI}. %\cite{Kumar2020JDI}
The objective of the work is to develop a simple technique to supervise and improve the performance of a compressive autoencoder, ensuring that most of the diagnosis-relevant data in whole slide histopathology images are retained in the compressed latent representation. It is worth noting that the main focus is on validating the theoretic feasibility of the proposed alternative techniques, rather than achieving similar or better results to the existing NIC and ROI-based compression approaches \cite{Chen2020JCO}.

%Now a day, all diagnostic services need to be integrated into the mobile and cloud-based medical ecosystem. This vastly interconnected and distributed network of devices and healthcare providers is often referred to as the Internet of Things (IoT). CPATH and AI-assisted diagnostic services also heavily rely on this network. Deep learning models are highly resourced intensive, and the data generation, training, and deployment of such models usually take place on different devices, often at distant locations also. This means that the individual sub-processes of CPATH need to be distributed on the IoT network \cite{Souza2020Sensor}. Advances in computation and digital technologies revolutionized healthcare in the past four decades. Improving quality and safety of recording and monitoring procedures, as well as speeding up diagnostic processes by aiding medical practitioners in their day-to-day work. Histopathology, in particular, has undergone major digitalization in recent years. Scanning whole histopathology slides and storing them in high-resolution Whole Slide Images (WSI) opened the door to a wealth of opportunities for pathologists \cite{Griffin2017DigitalPI}. 

Our research intends to develop and validate a simple but novel method for directing the attention of a special kind of neural network that can compress image data, called an autoencoder, to retain more of the meaningful information present in the input image. In the context of histopathological images, we target the compressed representation should contain more of the important features necessary for correct diagnosis when the autoencoder is supervised by a trained classifier network that is placed on top of it. The proposed supervision method does indeed promote beneficial changes in the autoencoderr’s encoding process, which in turn improves the accuracy of the label predictions made by the classifier.

%The remaining sections of the paper are organized as follows: The background techniques along with their existing works are studied in Section~\ref{literature}. The proposed methodology for compressing histopathological images with autoencoder is presented in Section~\ref{methodology} with the objectives and data preparation. The simulation environment, compressive autoencoder, classifier and ensemble methods are discussed in Section~\ref{simulation}. The simulation results analysis is carried out to fulfil the objective of the work and is presented in Section~\ref{results}. Finally, the outcomes of the proposed work and the future work are discussed in Section~\ref{conclusions}. 

\section{Background and literature review}
\label{literature}
%This section presents the background and literature review on digital pathology, image compression, and autoencoders. 
This section presents the background and literature review on image compression. 
When the compression rate of the lossless algorithm is not sufficient, it is necessary to allow for some information to be lost in the encoding process. Since the human eye can deal with a certain level of distortion or corruption in images, by removing this psycho-visual redundancy via quantization, much higher compression rates can be achieved. Hussain et al. \cite{hussain2018image} provide a good overview of the existing lossy compression techniques in their article. JPEG 2000 is the most widespread compression standard that allows for both lossless and lossy compression. A notable characteristic of JPEG 2000, and a major improvement over its predecessor algorithm (JPEG) that is worth mentioning here, is that it allows for user-controlled parameterization which enables different regions of images to be compressed at different rates \cite{Kanakatte2017EMBC}. Other commonly used evaluation metrics are Peak Signal-to-Noise Ratio (PSNR) and structural similarity (SSIM). Unfortunately, none of these metrics truly align with human evaluation of perceptual similarity \cite{Asiedu2022ARXIV}. 

In histopathology and medical imaging, the general standard is lossless compression, and it is preferred over lossy compression techniques so as not to jeopardise diagnostic performance \cite{Fan2021SR}. However, several studies have investigated the effects of lossy compression methods in healthcare and found that - depending on modality. More recently, Chen \cite{Chen2020JCO} in the field of computer assisted diagnostics, and studied the effects of JPEG2000 compression on the performance of deep learning models in detecting metastatic cancer cells in histopathological images. In the case of WSIs, a large percentage (around 80\%) of the slide area can be diagnostically irrelevant background or empty space, but compression artefacts in regions where tissue is present is highly undesirable \cite{Roy2022bioRxiv}. 

CNNs can learn to abstract semantic representations of the content in images, and they are the de facto tool of choice in computer vision problems. When trained in a supervised fashion, they can learn what information is important for the task at hand and what can be dismissed \cite{Mahendran2014arXiv}. %\cite{Dosovitskiy2015arXiv
Deep learning techniques have already been used in JPEG artefact removal with great results, replacing handcrafted model-based approaches \cite{Fu2019ICCV}.
%\cite{fu2021model}. 
In terms of using CNNs in the compression process itself, utilized CNN to flag regions of interest, which were in turn compressed with higher bit rate using a JPEG2000 encoder \cite{Prakash2016arXiv}. In recent studies, neural networks are taking over the compression process entirely \cite{Sadeeq2021iJOE}, %\cite{Bartler2021EUSIPCO}.
%\cite{Gao202ICCV}.
%\cite{Siwei2020IEEETrans}
Li and Ji \cite{Li2019arXiv} provide a great overview of the key concepts in neural image compression (NIC). In their article they make a keen observation about the general design and composition of NIC architectures; that given the successive encoding-decoding steps in compression tasks, they are all variations of a simple autoencoder design.

\section{Proposed Methodology}
\label{methodology}
In this section, we present the design of the autoencoder using supervised learning for the image compression. 
Autoencoder are neural networks comprised of an encoder bottom and a decoder top, and a bottleneck between them \cite{Alexandre2019arXiv}. They are usually symmetric in that the decoder mirrors the encoder part in its general structure, but this is not a necessary requirement. The key idea is that the output of an autoencoder is a reconstruction of the input and the network is optimized for minimizing the difference between them, i.e. the reconstruction error. 

Assuming unlabelled training inputs $x^{1},~ x^{2},~ . . . ,x^{n}$ and we can use it as in Eq. (1). The autoencoder inputs $x \in [0,1]^d$ for encoding with the hidden representation $y \in [0,1] ^d$ using mapping technique with a nonlinear function f as defined in Eq. (2).

\begin{equation}
    z^{i} = x^{i}
\end{equation}

\begin{equation}
    y = f(Wx + k)
\end{equation}

\begin{equation}
    p = f(W'y + k')
\end{equation}

\begin{figure}
	\centering
	\includegraphics[width=1\linewidth]{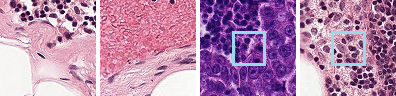}
	\caption{Sample Images from the Data Set}
	\label{fig:sample_images}
\end{figure}

\begin{figure}
	\centering
	\includegraphics[width=1\linewidth]{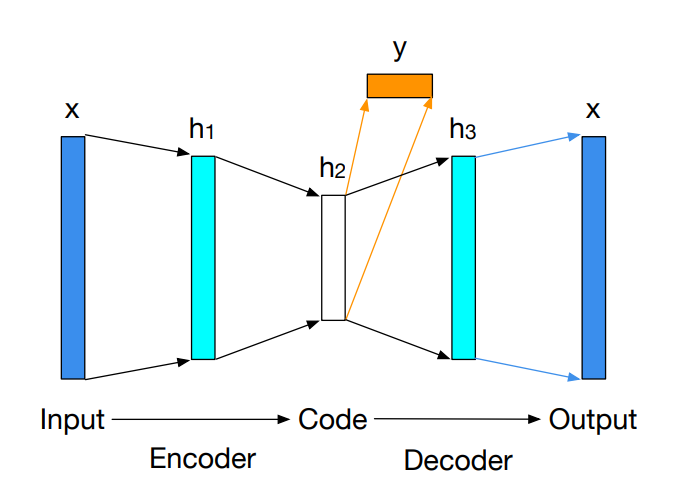}
	\caption{Autoencoder Structure }
	\label{fig:autoencoder_architecture}
\end{figure}

The basic structure of autoencoder for an image is shown in Figure~\ref{fig:autoencoder_architecture}. Since the bottleneck has fewer dimensions compared to the input and the output, the encoder  produces a compressed latent representation of the original image which makes this architecture a choice for compression problems.  This is in contrast with most neural networks that often expand on dimensionality when learning abstract semantic representations to solve supervised problems.

For image data, convolutional autoencoders are the most common architecture of choice, and the loss functions are usually MSE, root MSE (RMSE) or variations of the other similarity metrics discussed. Irrespective of the specific architecture design or particular loss function/distortion metric used, autoencoders will learn how to encode the input in a compressed latent representation and how to reconstruct the original image from that code by minimizing the difference between the input and output image. In case of convolutional autoencoder the weights are distributed among all the inputs and the reconstruction image $y$ can be defined using Eq. (4), where $c$ is the input bias per channel, $G$ is the set of latent features, and $W$ is flip operation computed between two weight dimensions.

\begin{equation}
    y = f(\sigma (g^{i}*W^{i} + c))
\end{equation}

\subsection{Supervised autoencoders}
The trouble with compressing images with autoencoders is that the process is unsupervised, although some times autoencoders are referred to as semi or self supervised models \cite{Chen2022arXiv}. A deep autoencoder optimizing on a MSE loss function, for example, is ultimately solving a regression problem (albeit in a non- linear fashion). And as such it is an unsupervised process, regardless of what semantic representation may or may not be present in the latent space. When it comes to compressing WSIs this way, the features retained by the autoencoder might be the best ones to minimize the average difference between the original and reconstructed images, but they might not contain the information needed for histopathological diagnosis, which calls for the need to be able to supervise autoencoders to make sure that they retain the necessary information? Le and White \cite{Bengio2018Book1} devised a Supervised Autoencoder (SAE) by adding a supervised loss (from label prediction) on the latent representation layer. By doing so, they managed to direct their autoencoder in representation learning towards features that were more important for the classification task.

In SAE, the supervised loss is included with the presentation layer, while in case of single hidden layer network it is added with the output layer. Assuming $k$ is the size of a single hidden layer, $F \in R^{d*k}$ represents the weights of the first layer, $W_{p} \in R^{k*m}$ stands for weights of output layer to predict y, weights $W_{r} \in R^{k*d}$  to reconstruct $x$, $L_{p}$ as supervised loss and $L_{r}$ as reconstruction error. The objective is defined in Eq. (5) \cite{Bengio2018Book1} as:

\begin{multline}
\frac{1}{t} \sum ^{t}_{i=1}[ L_{p}(W_{p}F_{x_{i}},x_{i})+L_{r}(W_{r}F_{x_{i}},x_{i})] = \frac{1}{2t}  \\
\sum ^{t}_{i=1}[ \parallel W_{p}E_{x_{i}}-y_{i} \parallel^{2}_{2}+ \parallel W_{r}E_{x_{i}}-y_{i} \parallel ^{2}_{2}]       
\end{multline}

The major challenge in image compression is information loss in the designed models. Researchers have proposed various loss functions depending upon the requirements and models. Let's assume supervised loss as $L_{s}$ with it related weight $W_{s},~ L_{r}$ as reconstruction error and its weight as $W_{r}$. We define the loss function $L(F)$ in Eq. (6) as:

\begin{equation}
L(F)= \frac{1}{t} \sum ^{t}_{i=1} L_{s}(W_{s}F{x_{i}},y_{p,i})+L_{r}(W_{r}F_{x_{i}},y_{r,i})
\end{equation}

Ultimately, the question if a reconstructed WSI after compression contains the necessary information or not boils down to whether the correct diagnosis can be made. However, visual evaluation of reconstructed images by actual histopathologists to assess the performance of a compressive autoencoder, especially during the development process, is simply unfeasible. More importantly, it isn’t really necessary after all. Consider a sample image (Figure~\ref{fig:sample_images}) from the Kaggle website, it describes: “A positive label indicates that the center $32x32px$ region of a patch contains at least one pixel of tumor tissue. Tumor tissue in the outer region of the patch does not influence the label. This outer region is provided to enable fully-convolutional models that do not use zero-padding, to ensure consistent behaviour when applied to a whole-slide image.”

\begin{figure}
	\centering
	\includegraphics[width=1\linewidth]{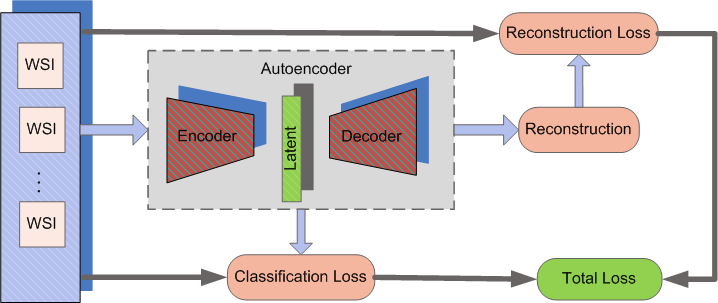}
	\caption{Presentation of the proposed model }
	\label{fig:Flow Diagram}
\end{figure}

As it has been noted, optimizing an autoencoder on the diagnostic performance and/or subjective evaluation of actual histopathologists is not a viable method. However, utilizing the next best thing, a well-trained classifier network to do the same is a rather reasonable approach that provides the basis for the chosen methodology.  Our proposed model is presented in Figure~\ref{fig:Flow Diagram}. The objective was to build an ensemble of neural networks that enables a compressive autoencoder in a supervised fashion to retain a denser and meaningful representation of the input histology images. 

The key steps of the proposed method were as follows:

\begin{enumerate}
    \item	Build and train a simple autoencoder to compress histopathological images.
   \item Train a classifier on the original, uncompressed images to correctly identify label/diagnosis.
   \item Build an ensemble with the trained autoencoder base and trained classifier top.
   \item Keep the classifier top untrainable and optimize the ensemble to correctly classify the reconstructed images.
    \item Since it is the only part that has trainable weights, the autoencoder should learn prioritize features important for classification (that the classifier top is looking for).
\end{enumerate}

One key advantage of using an autoencoder to compress images is that it can be broken up after being trained, and the encoding and decoding could be done independently and on separate devices. Which means that it is only the compressed representation of the original image that needs to be transmitted between them, thus reducing network traffic. 
%The type of autoencoder chosen for the ensemble was a simple compressive CNN. Although vibrational autoencoder seem to be in vogue for their probabilistic modelling approach and generative capabilities, reproducibility and interpretability are paramount in medical diagnostics. Therefore, the more straightforward design of a simple autoencoder was deemed more adequate. The network artefacts of simple convolutional autoencoders are well known, have a highly regular pattern and can be minimized with careful network design or could be later removed from the reconstructed images if need be. 

For developing the classifier, the most promising and reasonable approach was to use transfer learning, i.e. taking a publicly available pretrained model and adapt it to the task at hand. This method seemed more efficient and reliable than designing and training a network from scratch. 
%That being said, the resulting model did not need to achieve stunning results, but it had to perform reasonably well on the classification task to be able to supervise the Auto-encoder. Otherwise, a classifier that is merely” guessing” cannot provide any guidance for the Auto-encoder base as to what features to look for and to retain in the compressed latent representation. Similarly, an auto-encoder that already performs exceptionally well - produces high-quality reconstructed images and preserves almost all information from the input - has” no room to grow”. Meaning that virtually no feature selection is taking place, therefore there is no process to be influenced by the classifier top. Such an Auto-encoder would not be suitable for this project where the main objective is proof of concept: to validate the proposed ensemble method and its mechanism of action.

\section{Simulation}
\label{simulation}
\subsection{Environment set up}

The simulation is carried out in a Windows 10 operating system (with Intel i7 2.20GHz CPU, 8GB RAM and NVIDIA GeForce GTX 1050Ti GPU) in Python for Keras API and NVIDIA CUDA enabled TensorFlow-GPU as backend.
The data set chosen  is the PCam data set \cite{Bastiaan2018arXiv} curated for the Histopathologic Cancer Detection competition (Histopathologic Cancer Detection, 2018) and contained 220 025 labelled patches of lymph node sections for training purposes, as well as 57458 unlabeled patches for testing; in total close to 8 GB of data.

\subsection{Compressive autoencoder}
Several autoencoder architectures were built and tested with different compression ratios: reducing the dimensionality to 33\% (6x6x256), 66\% (12x12x128) and 83\% (12x12x160) of that of the original input images (96x96x3). As opposed to common practice in autoencoder design, the output of the encoder, i.e. the compressed code, was not flattened, as it was regarded as an unnecessary step. Although, it certainly could be done in the future if needed. The encoder and decoder parts were symmetric in all autoencoder.

\begin{figure}
	\centering
	\includegraphics[width=1\linewidth]{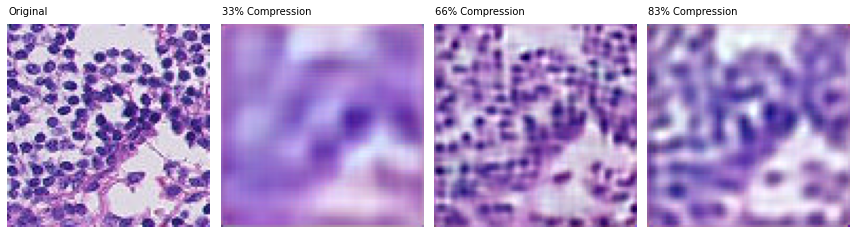}
	\caption{Comparison of the original and reconstructed images with different compression rates. }
	\label{fig:original_vs_reconstructed_images}
\end{figure}

They all followed a similar structure of having 3-4 blocks containing: (3-by-3) convolutional layers with ’same’ padding followed by batch normalization and ReLU activation function. In the encoder this was followed by (2-by-2) max pooling layers whereas in the decoder they were followed by (2-by-2) up-sampling instead. The autoencoder were trained for a 100 epochs, optimizing on MSE loss function as the measure of difference between the input and output images. By virtue of having such small models, the training never took longer than 15 minutes.

\begin{figure}
	\centering
	\includegraphics[width=0.7\linewidth]{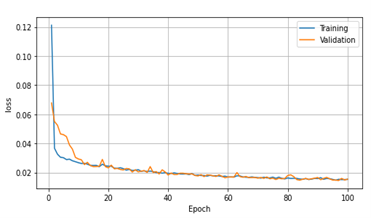}
	\caption{Autoencoder Summary}
	\label{fig:auoencoder_summary}
\end{figure}

Figure~\ref{fig:auoencoder_summary} depicts the performance evaluation and ultimately the selection of autoencoder models based on visual comparison of their respective output. Other major factor in selection of the right autoencoder to be used in the ensemble is influenced by the compression rate. Higher compression rates allow for more pronounced effects under the supervised training in terms of what features should be retained in the compressed representation. Thus, the autoencoder that reduced the dimensionality of the original images to 66\% was deemed as shown in figure 5.

\subsection{Classifier}
The method of choice to develop the classifier, as previously discussed, was transfer learning. Several models available from the Keras API were experimented with; e.g. EfficientNetB2, ResNet50, VGG16 and Xception that were already pre-trained on the ImageNet data set. The approach was used to adapt these models for binary classification by loading base model without the top layers and then adding a global average pooling layer and a dense layer with a single node using sigmoid activation function. But, all of the layers in the base models were set as non-trainable and as gradually training progressed, more and more layers were set trainable allowing the model to adapt to the specific computer vision and image classification problem which also referred as fine-tuning in transfer learning. However, validation loss is carefully monitored to avoid overfitting of the models during the training. Check- points/call-backs are often used for the models and saved in states during validation process to improve performance. 

Data augmentation is also deployed when required for rotating to flip the images either vertically or horizontally. The optimizer in all training was $Adam$, with a learning rate between 0.01-0.0001 as per situation requirement. The loss functions in all cases are binary cross-entropy. The classifier with the Xception base is selected to” supervise” the autoencoder in the ensemble with 0.95 accuracy and AUC-ROC score on the validation (uncompressed) data set. To illustrate the training history and performance of this model, graphs are shown in Figure 6.

\begin{figure}
	\centering
	\includegraphics[width=1\linewidth]{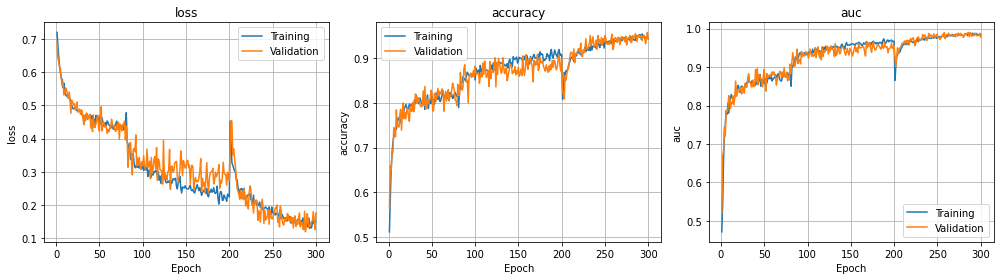}
	\caption{Classifier training with different performance measure}
	\label{fig:classifier_performances}
\end{figure}

In Figure~\ref{fig:classifier_performances}, the classifier is trained for 300 epochs in total and for the first 80 epochs only the classifier top is trainable. Then, last/top convolutional block of the exception base model is set trainable for the next 120 epochs during which the accuracy on the validation and training images started to diverge. For 176 000 training images and a batch size of 64 after 200th epoch data augmentation is implemented (rotation, horizontal/vertical flipping) to avoid overfitting along with training of weight adjustment. 

\subsection{Ensemble}

After a suitable compressive autoencoder and a classifier which correctly predict the target labels, or rather accurately enough, the procedure follows as:

\begin{itemize}
    \item Initially, set all the weights in the classifier untrainable.
    \item Add untrainable weight on top of the trained autoencoder.
    \item Train the whole assemble to accurately classify images.
\end{itemize}

Before the training of Xception classifier the optimizer and loss function are the same. Since, the weights of the classifier top cannot be trained thus the features used to predict the labels cannot be changed either or too. Moreover, the model is a kind of model which “knows what it’s looking for”. The AE will learn to priorities features that the classifier needs to make accurate diagnosis and retain them in the compressed latent code, something along the lines of a ROI-based compression method. The regions of interest are dictated by the supervising classifier.

\begin{figure}
    \centering
    \includegraphics[width=1\linewidth]{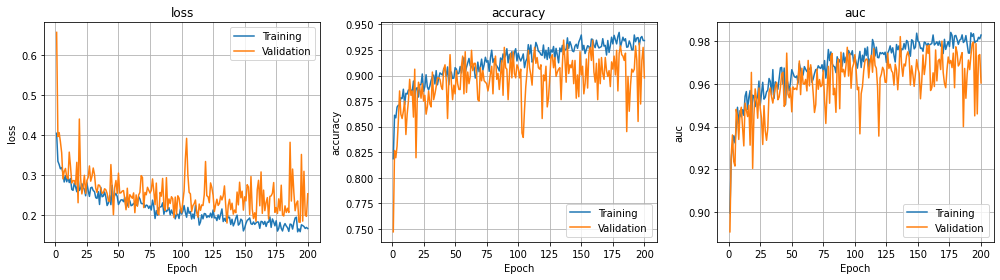}
    \caption{Ensemble training with different performance measure}
    \label{fig:ens_hist}
\end{figure}

Figure~\ref{fig:ens_hist} depicts the accuracy rate of the ensemble in the first epoch was quite low but gradually significant increment during the training, exceeding 0.9 by the 100th epoch. To avoid model state preservation when it has over-fitting during the training data, the evaluation loss and accuracy are closely monitored and checkpoints were created accordingly. The improvement in accuracy and AUC measures indicate the autoencoder is encoded so as to betterment the input images.

\section{Results}
\label{results}
\begin{figure}
    \centering
    \includegraphics[width=1\linewidth]{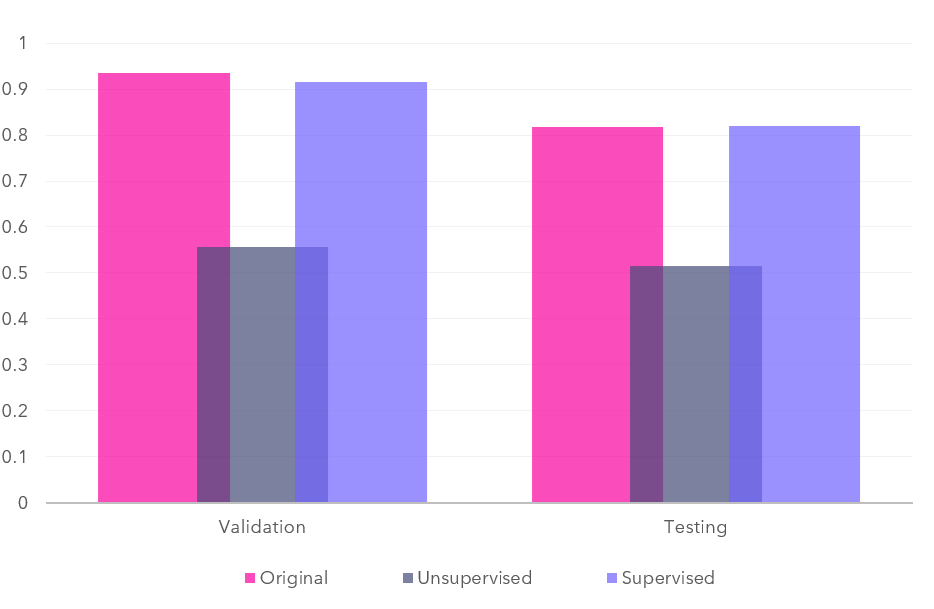}
    \caption{AUC-ROC scores of classifier performance}
    \label{fig:test_chart}
\end{figure}

It is being said that, 20\% percent of the training data is set aside for evaluation purposes during training and model development and in these cases accuracy and f1 scores are also utilized as represented in Figure~\ref{fig:autoencoder_summary}. Figure~\ref{fig:test_chart} shows the scores calculated based on the predictions made by the task-adapted Xception classifier on different images. The left side of the figure shows the scores obtained on images from validation data set, and on the right from test data set. The magenta bars show the AUC-ROC score for the label predictions of the original, uncompressed images. The grey bars indicate the classifier’s performance on the images reconstructed by the unsupervised autoencoder. The blue bars mark the classifier’s performance on the images reconstructed by the supervised autoencoder that has been trained in the ensemble.

%\begin{figure}
	%\centering
	%\includegraphics[width=1\linewidth]%{figures/AE_output.png}
	%\caption{Training history of the selected autoencoder}
	%\label{fig:training_history_autoencoder}
%\end{figure}

The classifier performs rather clumsily on the images which compressed and reconstructed by the unsupervised autoencoder, i.e. the simple compressive AE that has not been trained in the ensemble; with AUC-ROC scored just above 0.5 and that is close to the effectiveness of random guessing. In fact, it’s even worse than simply predicting the more prevalent negative diagnosis (approximately 60\% of the images have 0/negative labels) which signifies the unsupervised autoencoder is not very good at image compression.

\begin{figure}
    \centering
    \includegraphics[width=0.7\linewidth]{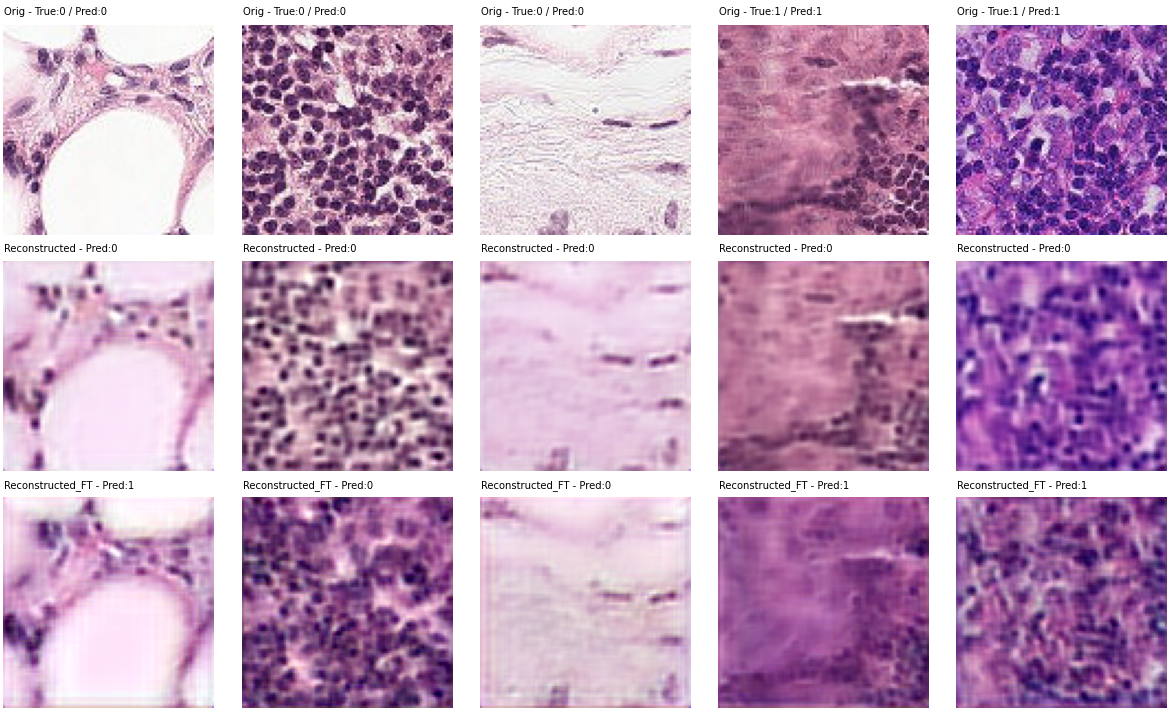}
    \caption{Sample images to illustrate the effects of compression with the unsupervised and supervised AEs}
    \label{fig:final_output}
\end{figure}

The resultant images of Figure~\ref{fig:final_output} indicate that not only is the supervising ensemble capable of influencing the autoencoder towards the features or regions of the images to retain in the latent representation, but that these features are actually influencing the reconstructive images. It certainly enables the classifier to overall improve more to accurate label predictions. In the Figure 10 the top row contains samples of the original, uncompressed images that are annotated with their true and predicted labels. The second or middle row shows the same images but after being compressed and reconstructed by the unsupervised AE, their labels predicted by the classifier. The bottom row contains the images (tagged with ’Reconstructed FT’ title) from the supervised autoencoder. 

\begin{figure}
	\centering
	\includegraphics[width=0.7\linewidth]{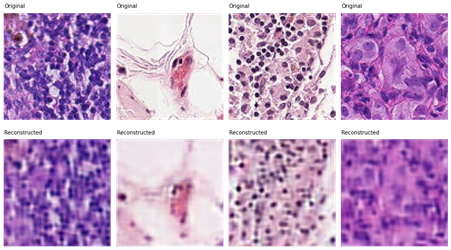}
	\caption{Examples of the images reconstructed by the selected autoencoder}
	\label{fig:reconstructed_images}
\end{figure}

To illustrate and assess the qualitative differences between the compression process and the reconstructed images with sample images have been provided in Figure~\ref{fig:reconstructed_images} using autoencoder. This could possibly indicate that the supervised autoencoder is prioritizing the spatial qualities and structural elements of the images, which are more informative for the label predictions, rather than overall colour. Once the image reconstructed by the supervised AE, the checkerboard network artefacts seem to be more pronounced in the peripheral regions and around the edges of the images, which could actually indicate a positive adaptation to the task, since, according to the data set description as it is only the center regions that inform the image labels. Similarly, these artefacts seem to be more pronounced in the white/empty regions of the images reconstructed by the supervised AE, compared to the unsupervised ones. This further supports the previous observation of the supervised AE placing more attention on regions important for the classification.

%Based on the findings described in the previous results, it has been demonstrated that the proposed methodology of supervising a compressive autoencoder by embedding it in an ensemble with a trained classifier top, is a viable approach for enriching the compressed latent representation with semantically meaningful information, and ultimately improving the quality of the compressed images. 

%Furthermore, the fact that we focus specifically on WSI compression and use histopathological data to demonstrate the viability of the proposed method suggests that this might be a viable compression technique in CAD and CPATH applications; fields that are in dire need for such solutions due to the exceptionally high volume of data.

\section{Conclusions and future work}
\label{conclusions}
%Although the work is not able to develop an out-of-the-box compression algorithm that is in any shape or form comparable to the performance of current image compression techniques, it does successfully meet the objectives that it is set out for. 

The main contributions of this work to the field of ROI-based neural compression is proof on concept; providing evidence for the viability of a new and simple supervision method for improving the performance of compressive neural networks. The exact specifications of the compression architecture may not be important; this method would possibly work for other compressive neural networks. Furthermore, if a diagnostic health- care provider already has a working, state-of-the-art classifier neural network, which they probably do, they now also have in their hands a quick and simple tool to improve upon their existing compression models and ultimately to decrease traffic in their IoT network.

%The main goal was to develop and validate a novel and simple method for supervising a compressive autoencoder, so that it learns to prioritise features that are more important to correctly classify histopathology images when choosing what to retain in the compressed feature space. It would be rather advantageous to be able to objectively measure as well as to visualise the structural differences between the images reconstructed by unsupervised and supervised CAEs. We could use Grad-CAM (Gradient-weighted Class Activation Mapping) technique or similar to localise and highlight image regions that are important for the label predictions. And focus on the changes in those parts of the reconstructed images. And if in the future this work could be successfully improved by the suggestions. Finally, it would be instrumental to apply this new compression method in a real life scenario; to actually develop a small mobile application based on this approach and to test the effectiveness of the encoding- transmission-decoding pipeline.

\section*{Declarations}

\vspace{-5 pt}

\begin{itemize}
\item \textbf{Ethical Approval} This research did not contain any studies involving
animal or human participants.
\item \textbf{Competing interest} The authors do not have any competing interest.

\item \textbf{Authors contribution} Agnes Barsi- Conceptualization \& Design, Simulation \& experiments, Prepared manuscript;
Suvendu Chandan Nayak - Prepared manuscript;
Sasmita Parida - Prepared manuscript;
Raj Mani Shukla - Conceptualization \& Design, Prepared manuscript
%All authors reviewed the manuscript
\item \textbf{Funding} There is no funding related to this paper

\item \textbf{Availability of data and materials} The data and materials will be avaiable on request

\end{itemize}

\bibliography{references}% common bib file
%\input output.bbl

\end{document}